\documentclass[twocolumn,a4paper]{article}
\usepackage{amssymb}
\usepackage{epsfig}  
\begin{document}
\twocolumn[%
\begin{center}
\textbf{\Large %
% TYPE THE TITLE OF YOUR ABSTRACT BELOW, BREAK LINES WITH \\ IF DESIRED
  Andreev scattering and conductance enhancement in mesoscopic
  semiconductor--superconductor junctions
% TYPE THE TITLE OF YOUR ABSTRACT ABOVE, BREAK LINES WITH \\ IF DESIRED
  }\vskip1ex {
% TYPE THE AUTHOR NAMES BELOW, HANDLE MULTIPLE AFFILIATIONS LIKE
% INDICATED IN THE SAMPLE ABSTRACT
  Niels Asger Mortensen$^1$, Antti-Pekka Jauho$^1$, Karsten
  Flensberg$^2$
% TYPE THE AUTHOR NAMES ABOVE, HANDLE MULTIPLE AFFILIATIONS LIKE
% INDICATED IN THE SAMPLE ABSTRACT
  } \vskip0.5ex
\textsl{ %
% TYPE THE AFFILIATION BELOW, HANDLE MULTIPLE AFFILIATIONS LIKE
% INDICATED IN THE SAMPLE ABSTRACT
  $^1$Mikroelektronik Centret, Technical University of Denmark,
  Building 345 east, DK-2800 Lyngby, Denmark\\
  $^2$\O rsted Laboratory, Niels Bohr Institute, University of
  Copenhagen, Universitetsparken 5, DK-2100 Copenhagen \O, Denmark
% TYPE THE AFFILIATION ABOVE, HANDLE MULTIPLE AFFILIATIONS LIKE
% INDICATED IN THE SAMPLE ABSTRACT
  }
\end{center}
]
%% START THE TEXT OF YOUR EXTENDED ABSTRACT HERE
Quantum transport in hybrid semiconductor--superconductor nanostructures has been shown to exhibit many of the mesoscopic effects
known from semiconductor systems \cite{BEENAKKER91a}, for instance
quantized conductance of the quantum point contact (QPC). Due to the
Andreev scattering \cite{ANDREEV64} these effects are modified and the
presence of a superconductor (S) opens up for studying new mesoscopic
phenomena, such as the quantized critical current in Josephson junctions
\cite{BEENAKKER91b}, but, also leads to a higher understanding of the
basic effects found in semiconductor structures.

An inherent difficulty in studying mesoscopic effects in
semiconductor--superconductor hybrid structures is the large Schottky
barrier which often forms at the interface. A large technological
effort has been invested in in improving the contact between the
superconductor and the two-dimensional electron gas (2DEG) of a
semiconductor heterostructure, and in recent years this has become
possible for e.g. GaAs-Al, GaAs-In, and InAs-Nb junctions. This
development motivates quantitative theoretical modeling of
sample-specific transport properties. The aim of our work is to model the conducting properties
of a ballistic 2DEG-S interfaces with a QPC in the normal region and
also to take into account scattering due to a weak Schottky barrier and/or
non-matching Fermi properties of the semiconductor and superconductor.

A theoretical framework is provided by the Bogoliubov--de Gennes (BdG)
formalism \cite{DEGENNES66} where the scattering states are
eigenfunctions of the BdG equation which is a Schr\"{o}dinger-like
equation in electron-hole space. The scattering approach to
phase-coherent dc transport in superconducting hybrids follows closely
the scattering theory developed for non-superconducting mesoscopic
structures. In zero magnetic field, Beenakker \cite{BEENAKKER92} found
that the Andreev approximation and the rigid boundary condition for
the pairing potential lead to a linear-response sub-gap conductance given by
\begin{equation}
\frac{G_{\scriptscriptstyle\rm NS}}{2G_0}
=  {\rm Tr}
\left(t t^\dagger \left[\hat{2} -tt^\dagger\right]^{-1} \right)^2=\sum_{n=1}^{N} \frac{T_n^2}{\left(2-T_n\right)^2}
\label{BEENAKKER}
\end{equation}
which, in contrast to the Landauer formula \cite{FISHERLEE81},
\begin{equation}
\frac{G_{\scriptscriptstyle\rm N}}{G_0}
=  {\rm Tr}\, t t^\dagger 
= \sum_{n=1}^{N} T_n,
\label{LANDAUER}
\end{equation}
is a non-linear function of the transmission eigenvalues $T_n$
($n=1,2,\ldots,N$) of $t t^\dagger$. Here $G_0=2e^2/h$ and $t$ is the $N\times N$
transmission matrix of the normal region, $N$ being the number of
propagating modes. The computational advantage of Eq.  (\ref{BEENAKKER}) over
the time-dependent BdG approach of De Raedt \emph{et al.}
\cite{RAEDT94} is that we only need to consider the time-independent
Schr\"{o}dinger equation with a potential which describes the disorder
in the normal region, so that we can use the techniques developed for
quantum transport in normal conducting mesoscopic structures.

\begin{figure}[hbt]
  \centering \epsfig{file=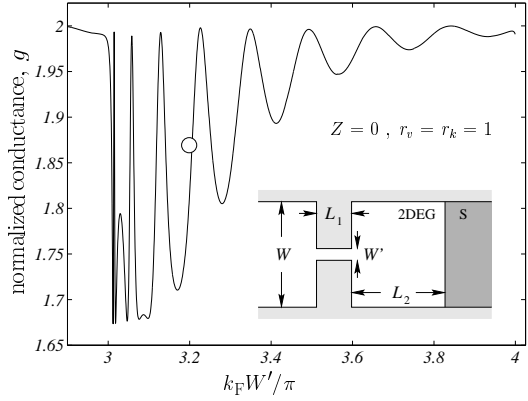,width=0.9\columnwidth}
\caption{Normalized conductance $g\equiv G_{\scriptscriptstyle\rm
    NS}/G_{\scriptscriptstyle\rm N}$ as a function of $k_{\rm
  F}W'/\pi$ for an ideal interface. The data-point ($\circ$) corresponds to the numerical result
  $(k_{\rm F}W'/\pi;g)=(3.2;1.87)$ of De Raedt \emph{et al.} \cite{RAEDT94}.}
\label{fig1}
\end{figure}

We study the geometry shown in the inset of Figure \ref{fig1} and
following recent work \cite{MORTENSEN99b} we model the QPC by a
wide-narrow-wide constriction \cite{SZAFER89}, the interface
scattering by a delta-function potential
\cite{BTK82,MORTENSEN99a}, and the transverse confinement by a
hard-wall confining potential. The scattering due to non-matching
Fermi velocities and Fermi momenta of the semiconductor and the
superconductor are taken into account by replacing the interface transmission
and reflection matrices of Ref. \cite{MORTENSEN99b} by

\begin{eqnarray}
\left(t_\delta\right)_{ww'}&=&\delta_{ww'}\frac{1}{\sqrt{\frac{\left[\Gamma(\theta_w)r_v+1\right]^2}{4\Gamma(\theta_w)r_v}}+i\frac{Z\sqrt{\Gamma(\theta_w)}}{\cos\theta_w}}\\
\left(r_\delta\right)_{ww'}&=&\delta_{ww'}\frac{\sqrt{\frac{\left[\Gamma(\theta_w)r_v-1\right]^2}{4\Gamma(\theta_w)r_v}}-i\frac{Z\sqrt{\Gamma(\theta_w)}}{\cos\theta_w}}{\sqrt{\frac{\left[\Gamma(\theta_w)r_v+1\right]^2}{4\Gamma(\theta_w)r_v}}+i\frac{Z\sqrt{\Gamma(\theta_w)}}{\cos\theta_w}}
\end{eqnarray}
where $r_v\equiv v_{\rm F}^{\scriptscriptstyle\rm (N)}/ v_{\rm
  F}^{\scriptscriptstyle\rm (S)}$ is the Fermi velocity ratio,
$r_k\equiv k_{\rm F}^{\scriptscriptstyle\rm (N)}/ k_{\rm
  F}^{\scriptscriptstyle\rm (S)}$ is the Fermi momentum ratio, and
$\Gamma (\theta )\equiv \cos \theta /\sqrt{\scriptstyle
  1-r_{k}^{2}\sin ^{2}\theta }$ \cite{MORTENSEN99a}.

We consider the device of Refs. \cite{RAEDT94,MORTENSEN99b}
with a relative width $W/W' =31.72$, an aspect ratio $L_1/W' = 5/1.6$,
and a relative length $L_2/W'=20/1.6$. Furthermore we consider a
junction between a GaAs 2DEG (in a GaAs-AlGaAs heterostructure) and a
superconducting Al film. For $T_{\rm F,GaAs}\simeq 100\,{\rm K}$,
appropriate parameters are given by $r_v=0.10$ and
$r_k=0.007$ \cite{MORTENSEN99a}.

\begin{figure}[hbt]
  \centering \epsfig{file=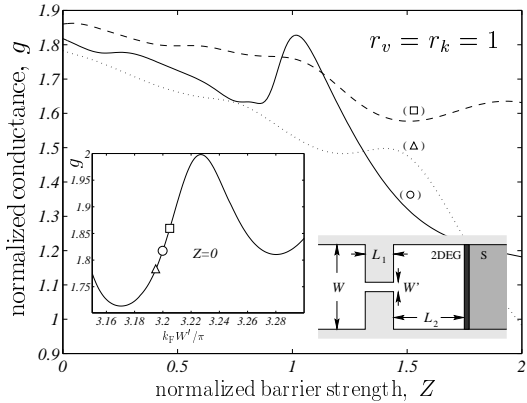,width=0.9\columnwidth}
\caption{Normalized conductance $g\equiv G_{\scriptscriptstyle\rm
    NS}/G_{\scriptscriptstyle\rm N}$  as a function of the normalized barrier strength $Z$
  for $k_{\rm F}W'/\pi = 3.195$ ($\triangle$), $k_{\rm F}W'/\pi = 3.2$
  ($\circ$), and $k_{\rm F}W'/\pi = 3.205$ ($\square$). The lower left
  insert shows $g$ as a function of $k_{\rm
    F}W'/\pi$ for $Z=0$.}
\label{fig2}
\end{figure}

Figure \ref{fig1} shows the normalized conductance $g\equiv
G_{\scriptscriptstyle\rm NS}/G_{\scriptscriptstyle\rm N}$ as a
function of $k_{\rm F}W'/\pi$ for an ideal interface. In Figure
\ref{fig2} we show the effect of a finite barrier at an interface with
matching Fermi properties. Compared to a similar system without a QPC,
see Fig. 2c of Ref. \cite{MORTENSEN99a}, the normalized conductance is
only weakly suppressed for low barrier scattering, $Z<1$, and only for
a very high barrier strength there is a cross-over from an excess conductance,
$g>1$, to a deficit conductance, $g<1$.

\begin{figure}[hbt]
  \centering \epsfig{file=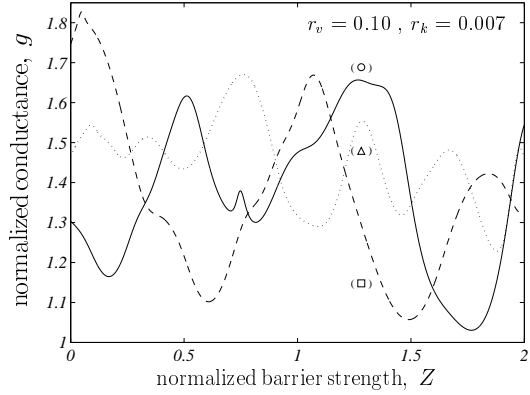,width=0.9\columnwidth}
\caption{Same plot as in Figure \ref{fig2}, but, with the
  non-matching Fermi properties taken into account.}
\label{fig3}
\end{figure}

In Figure \ref{fig3} we show how these results are modified when
taking the different Fermi properties into account. The detailed behavior is now changed but the overall weak effect of
the non-ideal interface on the normalized
conductance is the same. Comparing to a similar system without a QPC,
see Fig. 2a of Ref. \cite{MORTENSEN99a}, the normalized conductance
changes from $g\sim 0.2$ (at $Z=0$) to $g> 1.3$ in the presence of a
QPC in the normal region. 

In conclusion, the studied effect of a non-ideal interface with a Schottky
barrier and non-matching Fermi properties is very similar to the
reflectionless tunneling behavior in diffusively disordered junctions
\cite{BEENAKKER92,VANWEES92} where the net result is as if tunneling
through the barrier is reflectionless. In the case of a QPC instead of
a diffusive region, the presence of the QPC enhances the normalized conductance
even though there is a weak dependence on the barrier strength so that the tunneling is not perfectly reflectionless.

%% THE LAST LINE OF YOUR EXTENDED ABSTRACT SHOULD BE ABOVE THIS LINE
%% IF YOU HAVE REFERENCES, USE THE FOLLOWING thebibliography
%% ENVIRONMENT.
%% IF YOU DO NOT HAVE REFERENCES, COMMENT OUT (WITH A %) THE
%% FOLLOWING TWO LINES

\end{document}